\documentclass[10pt]{article}
\usepackage{amsmath,amsthm,latexsym,amssymb,amsfonts,epsfig,psfrag}

\usepackage{tikz}

\usepackage[top=2cm,bottom=3cm,left=3cm,right=3cm,a4paper]{geometry}

\def\be{\begin{equation}}
\def\ee{\end{equation}}
\def\ba{\begin{eqnarray}}
\def\ea{\end{eqnarray}}

\newcommand\nn{\nonumber}
\newcommand\q{\quad}

\renewcommand{\a}{\alpha}

\date{}

\begin{document}

\title{The continuum limit  of loop quantum gravity --\\ a framework for solving the theory}

\author{Bianca Dittrich
\\
\small 
 Perimeter Institute for Theoretical Physics, \\
 \small 31 Caroline Street North
  Waterloo, Ontario Canada N2L 2Y5 
}


\maketitle

\begin{abstract}
\noindent
The construction of a continuum limit for the dynamics of loop quantum gravity is unavoidable  to complete the theory. We explain that such a construction is equivalent to  obtaining the continuum physical Hilbert space, which encodes the solutions of the theory. We discuss iterative coarse graining methods to construct physical states in a truncation scheme and explain in which sense this scheme constructs a renormalization flow. We comment on the role of diffeomorphism symmetry as an indicator for the continuum limit.

\end{abstract}

\section{Solving the dynamics of loop quantum gravity}\label{intro}

Loop quantum gravity led to a rigorous non--perturbative framework, in which to formulate the dynamics of quantum gravity. It allowed fascinating insights into quantum geometry and a possible structure of quantum space time. To get a complete picture of the dynamics of the theory  -- in the form of constructing the so--called physical Hilbert space of wave functions satisfying the Wheeler deWitt equation -- we need to construct the continuum limit. In the framework presented here physical states, i.e.\ solutions of the Wheeler deWitt equations which give the equations of motions of the theory, are constructed by taking the refinement limit via a coarse graining procedure.

 The conceptual underpinnings of this framework rely on the inductive limit Hilbert space construction used in loop quantum gravity to define the continuum (so far kinematical) Hilbert space. We point out the powerful concept of this inductive limit construction if one allows for a generalization of the refinement maps that define the inductive limit Hilbert spaces.  It leads to a framework in which physical states are computed in a truncation scheme, where the type of truncation is determined by the dynamics itself. 

This procedure allows for an understanding of the dynamics of quantum gravity on all scales -- where a notion of scale  is given by the coarseness or fineness of configurations. The different scales of the theory are connected via the cylindrical consistency condition inherent in the inductive limit construction. This replaces the notion of renormalization flow in theories with a background scale. 

We start our considerations with a short explanation of the inductive limit construction in section \ref{inductive} and discuss the difference between kinematical and dynamical understanding of the continuum limit. In section \ref{dynamics} we start with the task to construct the physical Hilbert space of the theory and explain that it necessitates the construction of the refinement limit for the dynamics of the theory. This results in an iterative coarse graining scheme, in which physical states -- or amplitude maps -- are constructed in a certain truncation, labelled by the coarseness or fineness of the discrete structures involved. The relation of this scheme with a renormalization flow is clarified in section \ref{rflow}. Concrete realizations of this scheme in the form of (decorated) tensor network methods are shortly explained in section \ref{dtnws}. 
We then point out the powerful notion of diffeomorphism symmetry for discrete systems in section \ref{diff}. The realization of this diffeomorphism symmetry is necessary for the definition of physical states and also indicates that a continuum limit is reached.  In this sense physical states can only be defined in the continuum limit. We end with a discussion and outlook of future developments in section \ref{diss}.

\section{Continuum limit in canonical loop quantum gravity}\label{inductive}

Loop quantum gravity is formulated as a continuum theory, we therefore should clarify the need for a continuum limit in canonical loop quantum gravity. To this end we will shortly discuss how this continuum formulation is achieved (a complete discussion can be found in \cite{thomasbook}). The key point is to use a so called inductive limit construction for the kinematical Hilbert space of loop quantum gravity. Such an inductive limit construction needs the following ingredients
\begin{itemize}
\item[(1)] A directed partially ordered set of labels, in the case of the Ashtekar Lewandowski representation \cite{AL} given by a suitable set of graphs $\alpha$ embedded into the spatial manifold $M$. The partial ordering is induced by a set of refining operations (adding an edge, subdividing an edge, inverting an edge). 
\item[(2)] Hilbert spaces ${\mathcal H}_\alpha$ associated to these labels.
\item[(3)] Embedding maps $\iota_{\alpha\alpha'}:  {\mathcal H}_\alpha \rightarrow {\mathcal H}_{\alpha'}$ for each pair of labels with $\alpha \prec \alpha'$, i.e. $\alpha'$ is finer than $\alpha$. These embedding maps have to satisfy the consistency condition $\iota_{\alpha' \alpha''} \circ \iota_{\alpha\alpha'}=\iota_{\alpha\alpha''}$ for any triple $\alpha\prec\alpha' \prec\alpha''$. 
\end{itemize}
The inductive limit of Hilbert spaces is given by the 
\ba
{\cal H}&:=& \overline{ \cup_\alpha {\cal H}_\alpha \big{/} \sim}
\ea
where the equivalence relation is defined as follows: two elements $\psi_\alpha \in {\cal H}_\alpha$ and $\psi'_{\alpha'} \in {\cal H}_{\alpha'}$ are equivalent $\psi_\alpha \sim \psi'_{\alpha'}$ iff there exist a refinement $\alpha''$ of $\alpha$ and $\alpha'$ such that $\iota_{\alpha\alpha''}(\psi_\alpha)= \iota_{\alpha'\alpha''}(\psi'_{\alpha'})$. In words two elements are equivalent if they become equal under refinement eventually.

The inner product on the Hilbert spaces ${\cal H}_{\alpha}$ has to be compatible with this equivalence relation, that is {\it cylindrically consistent}
\ba\label{ccip}
\langle \psi_\alpha \,| \, \psi'_{\alpha}\rangle_\alpha &=& \langle \iota_{\a\a'}(\psi_\a) \,|\, \iota_{\a\a'}(\psi'_\a)\rangle_{\a'} \q .
\ea
Also observables, which are a priori given as family of observables ${\cal O}=\{{\cal O}_\a\}_\a$ defined on the Hilbert space ${\cal H}_\alpha$ have to be cylindrically consistent, that is
\ba\label{ccob}
\iota_{\a\a'}({\cal O}_\a \psi_\a) &=& {\cal O}_{a'} \iota_{\a\a'}(\psi_\a)  \q .
\ea
The conditions (\ref{ccip},\ref{ccob}) make the inner product and the observables well defined on the continuum Hilbert space, given by the inductive limit of the Hilbert space ${\cal H}_\alpha$. On a practical level they ensure that any calculation done on a given graph $\a$ (or any other discrete structure) gives the same result as on any refined graph.

Thus the construction of the inductive limit enables one to test the theory `along' discrete structures, such as the graphs $\alpha$. It is however {\it not} the case that the states are unknown away from the discrete structure in question. In fact  the embedding maps allow to reconstruct the states on an arbitrary refined graph $\alpha'$, starting from  states on a coarse graph $\alpha$. That is all additional degrees of freedom, associable to $\alpha'$ but not to $\alpha$ are being put into a specific state encoded in the embedding maps.  It is natural to  interpret this specific state as vacuum, in fact in the Ashtekar Lewandowski representation this state is given by the so--called Ashtekar Lewandowski vacuum. It is given as the (equivalence class represented by the) state associated to the empty graph $\alpha=\emptyset$ which carries a one--dimensional Hilbert space  ${\cal H}_\emptyset = \mathbb C$. The equivalence class of this vacuum state is characterized by the chosen embeddings -- turning this around the nature of the vacuum state characterizes the embeddings. 

The basic field variables of loop quantum gravity are given by the Ashtekar--Barbero connection $A^i_a$ and the triad densities $E^a_i$ \cite{newvar,barbero}. The connection is integrated and exponentiated to holonomies, along the edges given by the graph $\alpha$ the triads give rise to flux operators. 

The Ashtekar--Lewandowski vacuum is a totally squeezed state that gives maximal uncertainty to the connection and is maximally peaked at vanishing triad variables, that is formally
$\psi_{vac}( A) \equiv 1$.
Thus the states in any ${\cal H}_\alpha$ are highly distributional -- (spatial) geometry encoded in the triads is only excited along the graph $\alpha$. Away from this graph, all expectation values and fluctuations of the (smeared) triads are vanishing.

\subsection{Kinematical understanding of the continuum limit}

We thus come to our first (kinematical) understanding of a continuum limit in canonical loop quantum gravity. This is the construction of states in the Ashtekar--Lewandowski representation or even alternative representations  that can be interpreted as representing continuum geometries. Keeping the Ashtekar--Lewandowski (AL) representation the construction of coherent states has been explored \cite{coherent},  also coarse graining in the kinematical Hilbert space have been considered \cite{etera}. Here one however works with a fixed graph and therefore keeps the distributional nature of the states with respect to the excitations of spatial geometry  -- that is the states describe still a spatial metric that is  almost everywhere  totally degenerate.

It is also possible to construct alternative representation of the observable (holonomy -- flux) algebra of loop quantum gravity. The first such alternative representation \cite{koslowski} changes the vacuum from being peaked on a totally degenerate spatial geometry to one that is peaked on a non--degenerate (background) geometry. In this (Koslowski--Sahlmann) vacuum fluctuations of the triad are still vanishing. Note that the embedding maps for this Koslowski--Sahlmann representation are different from the one for the AL representations. As opposed to the AL representation the Koslowski--Sahlmann vacuum is not invariant under (spatial) diffeomorphisms anymore. 

Another alternative representation, that is based on a (space--time) diffeomorphism invariant vacuum has been recently proposed \cite{newvac} and can be understood as a dualization of the AL representation. The vacuum is now a totally squeezed states, that is peaked on flat connections, and maximally uncertain in the triad variables.  This vacuum is actually a physical state for BF theory, whose equation of motion demand vanishing curvature. This construction is based, as the AL representation on an inductive limit, however the label set is not given anymore by graphs but by triangulations.  The vertices (in $(2+1)$ dimensions) or edges (in $(3+1)$ dimensions) of this triangulation can support curvature excitations. Thus the states can be interpreted as piecewise flat geometries. (Note that in $(3+1)$ dimensions this flatness is with respect to he Ashtekar--Lewandowski representation, whereas in $(2+1)$ dimensions the flatness is with respect to the 3D spin connection.) In this sense this BF representation avoids a key problem of the AL representation, which is that AL states describe geometries which are almost everywhere totally degenerate. 

Thus, whereas the AL embeddings impose the vanishing of (`finer') triad operators, the BF embedding maps impose the vanishing of (`finer') curvature operators (built from holonomies). These embedding maps coincide with `naive time evolution maps' that arise in BF theory. In $(2+1)$ dimensions BF theory describes the dynamics of general relativity, and the BF vacuum defines therefore a physical state, giving rise to a physical Hilbert space.  This illustrates an important point -- namely that eventually the embedding maps should be chosen by the dynamics of the system. 

The BF representation has been also generalized  -- via a quantum group deformation of the underlying gauge group -- to a vacuum peaked on a homogeneously curved geometry \cite{DG16}.  In $(2+1)$ dimension the vacuum represents a physical state of general relativity with a cosmological constant, and is closely connected to the Turaev Viro state sum model. These different examples open up the questions of how many different quantum geometry realizations one is able to construct \cite{timeevol}.

A very different approach, which avoids the selection of a vacuum state, is being developed in \cite{Lanery}.  This framework replaces the inductive limit construction with a (dual) projective limit for the density functionals. But it is not clear yet, what kind of `typical states' result from this framework.

\section{Continuum limit for the dynamics of the theory}\label{dynamics}

We thus come to a second -- dynamical -- understanding of the continuum limit. This would be the construction of the continuum physical Hilbert space of states satisfying the Hamiltonian and diffeomorphism\footnote{Even if these constraints can be defined only a posteriori as discussed in section \ref{diff}.},constraints, that is the Wheeler DeWitt equations.  Such physical states are expected not to be normalizable with respect to the kinematical Hilbert space. In fact we have now at our disposal several kinematical Hilbert spaces, all based on an inductive limit, but with different embedding maps. 

We expect that also the physical Hilbert space can be organized in the form of an inductive limit Hilbert space. In this case the embedding maps $\iota_{\alpha,\alpha'}$ will  again differ from the embedding maps for the kinematical Hilbert space. We will outline here a construction of such a physical Hilbert space, which would then represent the continuum physical Hilbert space.

A strategy \cite{RAQ,thomasbook} to construct physical states, known as refined algebraic quantization, is   by `projecting' kinematical states via a so--called rigging map $\eta: {\cal D}_{kin}  \rightarrow {\cal D}_{phys}^*$.\footnote{Here ${\cal D}_{kin}$ is a dense subspace of the kinematical Hilbert space, ${\cal D}_{kin}\subset {\cal H}$, whereas ${\cal D}_{phys}^*$ is given by the algebraic dual of a dense subspace ${\cal D}_{phys}$ in ${\cal H}_{phys}$.} For totally constrained systems, where time evolution is a gauge transformation,  one can formally  write a `projector' onto the constraints as 
\ba\label{projector}
\left(\prod_I \delta(\hat C_I)\,\,  \psi \right) (X_{fin})  &=& \int {\cal D} N^I \, \exp\left( \frac{i}{\hbar} N^I \hat C_I \right) \psi  (X_{fin}) \nn\\ 
&=&  \int {\cal D} X_{ini} \int_{X_{ini},X_{fin} \, \text{fixed}}  \!\!\! \!\!\! {\cal D}X \exp  \left( \frac{i}{\hbar} S(X) \right) \,\, \psi_{ini}(X_{ini}) \, .\q\q
\ea 
In the second line we wrote  the path integral  over some set of configuration variables $X$ with the corresponding action $S(X)$ for general relativity. Equation (\ref{projector}) states that this path integral  serves as a (formal) projector onto states satisfying the Hamiltonian and diffeomorphism constraints \cite{proj}. 

The path integral is however only a formal object -- so far the only way to make it well defined is to turn to a discretization. This is one route to spin foam models \cite{foams}, for which the (boundary) variables $X$ can be made to match those of loop quantum gravity.\footnote{More precisely  the boundary Hilbert spaces match \cite{projspinnetworks}, at least on the discrete level.} A discretization comes however with several drawbacks:
\begin{itemize}
\item[(a)] A discretization typically breaks diffeomorphism symmetry for 4D gravity theories \cite{bahrdittrich09a}. This prevents the discrete path integral to be a projector onto constraints,  these are rather weakened to pseudo constraints \cite{consistent,bahrdittrich09a,hoehn1}.
\item[(b)]  Related to the loss of diffeomorphism symmetry the path integral will in general depend on the choice of discrete structure, i.e.\ choice of  (bulk) triangulation. This gives the triangulation an unwanted physical significance.
\item[(c)] There are many classical and quantum ambiguities in constructing the discrete amplitudes.
\item[(d)]  The discrete path integral (\ref{projector}) can be defined on the Hilbert spaces ${\cal H}_{\alpha}$ associated to a given discretization $\alpha$. However as an operator on the family of Hilbert spaces  ${\cal H}_{\alpha}$ the path integral will in general not be cylindrically consistent and thus not be well defined on the continuum Hilbert space ${\cal H}$ \cite{bahrproc,bd12b,bahr14}.  
\item[(e)] Finally a discrete path integral requires also an organization of the target (physical Hilbert) space as an inductive limit. The path integral as an operator should then also be cylindrically consistent with respect  to dynamical  embedding maps \cite{bd12b,timeevol} describing the physical Hilbert space. 
\end{itemize}

We will argue that all these issues can be addressed by  coarse graining the initial discrete path integral. As we will see this can also be interpreted as refining and amounts to the construction of  the continuum limit for the discretized path integral.

To achieve the continuum limit for the dynamics of quantum gravity means in particular  to turn the path integral into a cylindrical consistent operator, that is solve issues (d) and (e). We will describe here an iterative coarse graining process that aims at the construction of such a cylindrical consistent path integral. 

This iterative process produces a coarse graining flow. Fixed points of such coarse graining flows often enjoy an enhanced symmetry. Several examples \cite{bahrdittrich09b,steinhaus11} and arguments (see section \ref{diff}) show that in particular diffeomorphism symmetry is likely to be restored, which addresses problem (a). The same examples and the realization of diffeomorphism symmetry in the discrete as so called vertex displacements show that diffeomorphism symmetry is equivalent with triangulation independence  \cite{dittrichreview,steinhaus11}, which resolves problem (b). Finally the coarse graining flow is considered on a space of models. Such a flow allows the characterization of relevant and irrelevant directions in this space of models, which addresses the issue (c). In particular, diffeomorphism invariance and triangulation independence are extremely strong requirements, thus one can hope that a discrete model satisfying these requirement (and leading to a suitable semi--classical limit) is, if it exist at all, unique.  

We will furthermore argue that the issue (e) will lead to 
\begin{itemize}
\item[(f)] a  notion of physical vacuum for quantum gravity.
\end{itemize}

This physical vacuum will be encoded into amplitude maps ${\cal A}_\alpha :{\cal H}_\alpha \rightarrow {\mathbb C}$. These maps define the amplitudes for the cylindrically consistent path integral and thus replace the `bare' amplitudes of the initial discretization of the path integral. Here the label $\alpha$ stands for a discretization that can be obtained by refinement from an `empty' discretization  $\emptyset$  with ${\cal H}_\emptyset={\mathbb C}$.

The amplitude map applied to a  boundary wave function\footnote{The framework \cite{oeckl} introduces a generalization of Cauchy boundaries to boundaries of arbitrary regions, which is useful in this context.}  $\psi_\alpha \in {\cal H}_\alpha$ gives the pairing of this wave function with the wave function ${\mathbf K}_{\emptyset \alpha} \psi_\emptyset$, resulting from a refining time evolution of the (kinematical no boundary) wave function $\psi_\emptyset$ to a wave function associated to the boundary $\alpha$. This refining time evolution is given by a path integral and can therefore be understood to implement a rigging map, see equ. (\ref{projector}). That is we consider
\ba\label{defam}
{\cal A}_\alpha(\psi_\alpha)&:=&  \int {\cal D} X {\cal D} X_\alpha \exp \left(-\frac{i}{\hbar} S (X,X_\alpha) \right) \,\psi_\alpha (X_\alpha) \nn\\
&=& \langle \psi_\emptyset  | ({\mathbf K}_{\emptyset \alpha})^\dagger | \psi_\alpha \rangle \,=\,  \eta(\psi_\emptyset) \cdot \psi_\alpha \,=:\, \langle \psi_\emptyset  | \psi_\alpha \rangle_{phys} \q .
\ea
Here we wrote the path integral\footnote{We wrote complex conjugated path integral  amplitudes $\exp(-  \frac{i}{\hbar} S (X,X_\alpha) )$ to indicate the complex conjugation of the wave function evolved from $\psi_\emptyset$. In spin foams (and in other approaches which incorporate in (\ref{projector}) an integration over positive and negative lapse) the sum over the basic variables includes a sum over orientations of space time. This leads to real amplitudes. This feature is important to obtain the projector property of the path integral. } with bulk variables denoted by $X$ and  boundary variables denoted by $X_\alpha$. We have to regularize this path integral on a discretization, that fits in-between the two boundaries $\emptyset$ and $\alpha$. This discretization  introduces of course the problems mentioned above, turning the expressions in the second line into not well defined ones. We will discuss below an iterative procedure to take the refinement limit of this discretization, that addresses these problems. 

We discussed in (\ref{projector}) that the time evolution operator in the form of the path integral should act as a projector onto physical states, which defines the rigging map $\eta$, here applied to the no--boundary (kinematical) wave function $\psi_\emptyset$. The last equation just displays the definition of a physical inner product in the refined algebraic quantization procedure \cite{RAQ}
\ba
\langle \psi_{\alpha} \,|\psi'_{\alpha'} \rangle_{phys} \,:=\, \eta(\psi_\alpha)\cdot  \psi'_{\alpha'}
\ea
between the projections of two kinematical states $\psi_\alpha$ and  $\psi'_{\alpha'}$. (It suffices to apply the rigging map once, as it is given by a time evolution which acts as an usually improper projector.)

The amplitude maps encode the dynamics of the system \cite{PerezRovelli} and will replace the `bare' amplitudes of the initial discretized path integral.  Note that such a discretized path integral is often built by associating amplitudes ${\cal A}_B$ to basic building blocks $B$. Indeed from the definition (\ref{defam}) the basic amplitudes ${\cal A}_B$ give the  amplitude map  in the coarsest triangulation possible. To this end we assume that one can refine the empty discretization $\emptyset$ to the one given by the boundary of $B$ by gluing the building block $B$ to $\emptyset$.

The iterative refinement process will replace these basic amplitudes with improved amplitudes ${\cal A}_\alpha$ by (i) refining the bulk discretization and (ii) also allowing a refining of the boundary discretization, that is generalize from the boundary of $B$ to finer boundary discretizations $\alpha$. 
This generalization of the basic building blocks, that allows the incorporation of more boundary data, is important to convert non--local couplings, that inadvertently are produced by coarse graining to local (nearest neighbour) couplings of the improved amplitudes.

The end point of the construction should lead to a cylindrically consistent amplitude map satisfying cylindrical consistency
\ba
{\cal A}_{\alpha'}(\iota_{\alpha\alpha'}(\psi_\alpha))&=& {\cal A}_\alpha(\psi_\alpha)
\ea
with respect to certain embedding maps $\iota_{\alpha\alpha'}$. As we will argue below, it might be much easier to construct such cylindrical consistent amplitudes if we replace the kinematical embedding maps with dynamical ones.

Such cylindrical consistent amplitude maps are then defined on a continuum Hilbert space ${\cal H}_{[ \emptyset]}$ associated to the equivalence class of discretizations, that can be obtained by applying refinement operations to the empty discretization $\emptyset$.

This brings us to the second interpretation of the amplitude maps as representing the (dualized) physical vacuum.  This interpretation is due to two points: 

Firstly we defined the amplitude map via a refining time evolution starting from a `no--boundary' discretization $\emptyset$. The resulting wave function can be seen as a Hartle Hawking no--boundary wave function \cite{hhv}.\footnote{The actual proposal \cite{hhv} Wick rotates part of the time evolution. We do not assume such a Wick rotation here, which would indeed be hard to define in a completely background independent context.}  This point is also strengthened as the amplitude map ${\cal A}_\alpha(\cdot)=\eta(\psi_\emptyset)\cdot$ results from applying the rigging map, to the kinematical vacuum $\psi_\emptyset \in {\mathbb C}$, which one would expect to carry the notion  of having no excitations and leading to a homogeneous state, see also \cite{florian}. 
This concept of generating a vacuum state by refining time evolution comes also up in formulations incorporating evolving phase spaces \cite{hoehn2} or Hilbert spaces \cite{hoehnq}, classical and quantum examples that support this interpretation can be found in \cite{timeevol}. In the formulation employed here evolving Hilbert spaces are taken into account via the concept of inductive limit Hilbert spaces.

Secondly, we will use the amplitude maps to define  dynamical embedding maps. That is the amplitude maps lead to an improved, and in the refinement limit, perfect discretization of the path integral. This path integral can be used to define a refining time evolution, interpolating between a boundary $\alpha$ and a refined boundary $\alpha'$. However, as we discussed, there is no proper time evolution in diffeomorphism invariant systems, it rather acts as a projector onto physical states. In case the initial state $\psi_\alpha$ is physical, the resulting state $\psi_{\alpha'}$ should therefore be equivalent to $\psi_\alpha$. This is realized if we assume an inductive limit structure for the physical Hilbert space and use the refining time evolution  as (dynamical) embedding maps $\iota_{\alpha\alpha'}={\mathbf K}_{\alpha\alpha'}$, as proposed in \cite{timeevol}. 

Note that such embedding maps have to satisfy the consistency conditions $\iota_{\alpha' \alpha''} \circ \iota_{\alpha\alpha'}=\iota_{\alpha\alpha''}$ for any triple $\alpha\prec\alpha' \prec\alpha''$, as discussed in section \ref{inductive}. For a (refining) time evolution these conditions follow  from Kuchar's requirement of a path independence of evolution \cite{kuchar}, which is equivalent to the constraint algebra being consistent, that is first class, which itself signifies that diffeomorphism symmetry is correctly implemented. We can therefore expect this consistency condition to hold in the refinement limit, in which we hope to restore diffeomorphism symmetry.

Another aspect of path independence of evolution is a condition involving as an in--between state one that is finer than the final state:
\ba\label{cg}
{\mathbf K}_{\alpha'' \alpha'} \circ  {\mathbf K}_{\alpha\alpha''}\,=\, {\mathbf K}_{\alpha \alpha'}
\ea 
for $\alpha \prec \alpha'\prec \alpha''$. If in addition we can identify ${\mathbf K}_{\alpha\alpha'} = ({\mathbf K}_{\alpha' \alpha})^\dagger$, which should hold due to the projector property of time evolution, it follows that  the amplitude maps are cylindrically consistent for dynamical embedding maps $\iota_{\alpha\alpha'}={\mathbf K}_{\alpha\alpha'}$:
\ba\label{iter1}
{\cal A}_{\alpha'}(\iota_{\alpha\alpha'} \psi_\alpha)&=& \langle \psi_\emptyset  | ({\mathbf K}_{\emptyset \alpha'})^\dagger  | {\mathbf K}_{\alpha\alpha'} \psi_\alpha \rangle  \nn\\ \  &\stackrel{ (\ref{cg})} {=}&
\langle \psi_\emptyset  | ({\mathbf K}_{\emptyset \alpha})^\dagger   | \psi_\alpha \rangle 
\q\q\q\,=\, {\cal A}_{\alpha}(\psi_\alpha) \q .
\ea
This suggest to also change the embedding maps on the kinematical Hilbert space, as this simplifies the construction of a cylindrical consistent amplitude map. 

Indeed we can take (\ref{iter1}) as defining an iterative procedure to improve the amplitude maps, in particular regarding property (\ref{cg}). To this end  we understand the term on the RHS of the first line in (\ref{iter1}) as consisting of two steps. 
The first is the computation of  $\langle \psi_\emptyset  | ({\mathbf K}_{\emptyset \alpha'})^\dagger$, that is the basically the amplitude functional ${\cal A}_{\alpha'}$ for a more refined boundary $\alpha$. One would build such an amplitude functional from gluing amplitudes ${\cal A}_\alpha$ for less refined boundaries $\alpha$. 

As we want to define an iterative process that improves the amplitude maps ${\cal A}_\alpha$, we need to find a way to  `evolve back' the amplitudes ${\cal A}_{\alpha'}$ to the boundary Hilbert space ${\cal H}_\alpha$, which is done by using the dynamical embedding map $\iota_{\alpha\alpha'}={\mathbf K}_{\alpha\alpha'}$. Thus one defines the improved amplitudes ${\cal A}^{imp}_\alpha$ as
\ba\label{imp}
{\cal A}_{\alpha}^{imp} &=&  \langle \psi_\emptyset  | ({\mathbf K}_{\emptyset \alpha'})^\dagger  | {\mathbf K}_{\alpha\alpha'} \psi_\alpha \rangle \q .
\ea
Here both $({\mathbf K}_{\emptyset \alpha'})^\dagger $ and ${\mathbf K}_{\alpha \alpha'}$ are built from using the initial ${\cal A}_\alpha$ as basic amplitudes. 

The process is repeated for the improved amplitudes ${\cal A}_{\alpha}^{imp}$ until the procedure converges to a fixed point ${\cal A}_{\alpha}^{fix}$. This fixed point amplitude can be used to proceed to a more refined pair of boundaries $(\alpha',\alpha'')$ with $\alpha' \prec \alpha''$ to find the next fixed point amplitude ${\cal A}^{fix}_{\alpha'}$ and so on.

One can take this amplitude ${\cal A}^{fix}_{\alpha'}$ and aim to construct a dynamical embedding map $\iota_{\alpha\alpha'}={\mathbf K}_{\alpha \alpha'}$ from a coarser boundary $\alpha$ to  a finer one $\alpha'$. This allows to consider the pull back ${\cal A}_{\alpha}^{fix,\alpha'}:= \iota_{\alpha\alpha'}^*{\cal A}^{fix}_{\alpha'}$. This amplitude will differ from ${\cal A}_{\alpha}^{fix}$, the amplitude constructed taking less boundary data, namely the pair $(\alpha,\alpha')$ into account. Because of this we see ${\cal A}_{\alpha}^{fix,\alpha'}$ as an improvement on ${\cal A}_{\alpha}^{fix}$. Iterating in this way one constructs amplitude maps that are satisfying the cylindrical consistency conditions for finer and finer boundaries.

Tensor network renormalization schemes make this procedure explicit, by specifying more in detail how to construct the refined amplitudes ${\cal A}_{\alpha'}$ and the embedding maps $\iota_{\alpha\alpha'}\sim {\mathbf K}_{\alpha\alpha'}$ from the amplitudes ${\cal A}_\alpha$ for coarser boundaries $\alpha$. We will explain a tensor network algorithm in section \ref{dtnws}.

Once one has constructed amplitude maps that are cylindrically consistent (to a satisfying degree), one can use these amplitude maps to define an improved discretization of the path integral (\ref{projector}) and with it the rigging map.   This is using the interpretation of the amplitude maps as giving the amplitudes of building blocks, which can now carry more boundary data. 

Let us examine the gluing properties of these improved building blocks, in particular in which sense the amplitude for a given (finer) boundary can be obtained from gluing building blocks with coarser boundary. If this would be the case we would achieve independence from the chosen discretization, i.e. \ form the decomposition of a given region into building blocks.

For this consider a simplified situation with two manifolds of topology $\Sigma \times [0,1]$ glued along a common $\Sigma$ hypersurface. The amplitude for the first manifold with boundaries $\alpha,\alpha'$ is given by ${\bf K}_{\alpha\alpha'}$, for the second manifold we have ${\mathbf K}_{\alpha'\alpha''}$ so that the glued amplitude is
\ba\label{glue1}
{\mathbf K}_{\alpha'\alpha''}\circ {\mathbf K}_{\alpha\alpha'} \, \stackrel{?}{=} \, {\mathbf K}_{\alpha\alpha''} \q .
\ea
Thus discretization independence (here invariance under subdivision) would be realized if the equality in (\ref{glue1}) holds.  This equation can however not be true for arbitrary coarse in-between boundary $\alpha'$. A very coarse $\alpha'$ would restrict the amount of information that can propagate from $\alpha$ to $\alpha''$.\footnote{The equality can hold however in topological theories which do not have local propagating degrees of freedom.} Thus $\alpha'$ should in general be finer than both $\alpha$ and $\alpha''$. In this case equation (\ref{glue1}) coincides with (\ref{cg}) (or its time reversal), and thus is expected to hold for cylindrically consistent amplitudes.

The situation is less clear--cut if we generalize to  situations where only certain parts of the boundary are glued. However, as this is also used in the coarse graining procedure which builds such cylindrical consistent amplitudes one would expect that -- depending on the coarseness of the outer boundaries not glued over -- the gluing property is satisfied  to better and better degree for finer and finer boundaries and in particular satisfied exactly if one takes for the boundary glued over the refinement limit. For subtleties that come up even in the continuum, see \cite{oeckl}.

\section{ Renormalization flow and scale in background independent theories} \label{rflow}

Here we want to discuss the relations and differences of the framework developed in section \ref{dynamics}, where the construction of cylindrical consistent amplitudes is central, to the understanding of renormalization flow in systems with a notion of background (scale)  \cite{oecklr,zapata,bahr14}. We will in particular provide an extension of aspects developed in the work  \cite{bahr14} from the AL embedding maps to dynamical embedding maps.

Consider a system with discretization scale $a'$, whose dynamics is defined by amplitudes ${\cal A}^{a'}(X')$ (e.g. $ \exp(\frac{i}{\hbar} S(X'))$) , depending on variables $X'$ (defined at scale $a'$). 
The Wilsonian renormalization flow \cite{wilson} defines effective amplitudes ${\cal A}^{a}$ at a larger scale $a$ through the condition
\ba\label{flow}
\int_{B^{a',a}_{X}(X')=X}  {\cal D} {X'}\,{\cal A}^{a'}(X') \,  &=& \int  {\cal D} X\,{\cal A}^{a}(X) \, \q .
\ea
Here we denote by  $B^{a',a}_X$  a blocking function that determines how the microscopic degrees of freedom $X'$ are coarse grained into the coarser variables $X$. 

Repeating (\ref{flow}) at different pairs of scales will give a renormalization flow of the amplitudes ${\cal A}^a$ parametrized by the scale $a$. The amplitudes at coarser scales encode the `effective' dynamics of the system and allow to determine the expectation values of sufficiently coarse observables ( that can be expressed in the variables $X$ of this scale): 
\ba\label{obsc}
\int {\cal O}(   B^{a',a}_X(X') ) \, {\cal A}^{a'}(X')  {\cal D} X' &=& \int {\cal O}(X)\,  {\cal A}^{a}(X) {\cal D} X \q .
\ea

In background independent systems we do not have a background scale available. Instead there are two entities, which replace the background scale: one is the discretization labels $\alpha$ characterizing the coarse-- or fineness of a given boundary. The other is the geometry, which is part of the dynamical variables and thus determined by the boundary data or wave function.

The renormalization trajectory ${\cal A}^a$, parametrized by the scale $a$ is replaced in background independent systems by the cylindrically consistent family of amplitude maps. Thus a cylindrically consistent family of amplitude maps defines a renormalization flow. 

To see this consider the path integral over a certain region built from building blocks $B$ or regions with a certain homogeneous boundary fineness $\alpha$. Subdivide each of these building blocks into further building blocks $\{B'\}$. We then want to compare the path integral based on amplitude maps for building blocks $B$ with the path integral based on amplitude maps for building blocks $B'$. Here the amplitudes ${\cal A}_B$ for a building block $B$ are defined from the (cylindrical consistent) amplitude maps via
\ba
{\cal A}_{\alpha(B)}(\psi_{\alpha(B)}) &=& \int  \overline{{\cal A}_{B} (X)} \, \psi_{\alpha(B)} (X)
\ea
where $\alpha(B)$ denotes the boundary of the building block $B$.  

Similarly we define a kernel $\iota_{\alpha,\alpha'}(X_{\alpha'}, X_{\alpha})$ for the embedding maps $\iota_{\alpha,\alpha'}$ by
\ba
\iota_{\alpha,\alpha'}(\psi_\alpha)\,(X_{\alpha'}) &=& \int {\cal D} X_\alpha \,\,\, \, \iota_{\alpha,\alpha'}(X_{\alpha'},X_{\alpha})\, \psi_\alpha(X_\alpha)
\ea
where $X_\alpha$ denote the boundary variables of $\alpha$ \footnote{e.g. if one understands the Hilbert space ${\cal H}_\alpha =L_2(  {\cal C}_\alpha, {\cal D} X_\alpha ))$ with ${\cal C}_\alpha$ denoting the configuration space and ${\cal D}X_\alpha$ the measure.} and $X_{\alpha'}$ those of $\alpha'$.

To connect the amplitudes for $B'$ and $B$ we integrate over the shared boundary variables when gluing the building blocks $\{B'\}$ to $B$. This will however result into a finer boundary than for the original building block $B$. We thus need to use  embedding maps $\iota_{\alpha, \alpha'}$ from the boundary $\alpha=\alpha(B)$ of $B$ to the boundary $\alpha'=\alpha'(\{B'\})$ of the set of glued building blocks $\{B'\}$. These embedding maps are applied in the inverse direction, as these act indeed on the boundary wave function, with which the amplitude is paired.

We denote by $\alpha''$ the boundary of a given building block $B'$ and with $\cup \alpha''/ \alpha'$ the inner (shared) boundaries in the gluing of the set $\{ B'\}$ to $ B$. The amplitude $\tilde {\cal A}_{ B}$ constructed from the ${\cal A}_{B'}$ is then given as
\ba\label{eff}
\tilde {\cal A}_B(X_\alpha) &=& \int  {\cal D} X_{\alpha'}  \left( \int   {\cal D} X_{\cup \alpha''/ \alpha'}  \prod {\cal A}_{B'} (X_\alpha')  \right) \overline{\iota_{\alpha\alpha'}(X_\alpha,X_{\alpha'})} \q .
\ea
The arguments from the previous section show that at least approximately we can expect ${\cal A}_B =\tilde {\cal A}_B$. 
Thus the ${\cal A}_B$ are indeed (also) effective amplitudes, that is they can be obtained by integrating out degrees of freedom starting from amplitudes ${\cal A}_{B'}$. A fixed point condition follows if we choose the building blocks such that $B=B'$.

Comparing with the definition of the Wilsonian renormalization flow (\ref{flow}) we can argue that the role of the pair of scales $(a',a)$ there is taking over by $(\alpha',\alpha)$. (As the original building blocks $B'$ might have the same boundary as the effective building blocks $B$, that is $\alpha''=\alpha$, we rather compare with the boundary of the set of glued building blocks $\{B'\}$.) 
The blocking functions $B^{a',a}$ are replaced by the embedding maps $\iota_{\alpha,\alpha'}$, which allow for more general constructions. A replacement of the (bulk) observable condition (\ref{obsc}) can also be stated \cite{bahr14} and is equivalent to the cylindrical consistency condition for (boundary) observables (\ref{ccob}).

Here we argued from the `boundary' cylindrical consistency of the amplitude maps on the boundary Hilbert space towards a `bulk' cylindrical consistency of the path integral measure (which we here understand to include  the amplitudes ${\cal A}_B$). This last point is the starting point of \cite{bahrproc,bahr14} for configuration spaces of connections and with the AL embedding maps. See also the discussion in \cite{bahr14} for a derivation of boundary cylindrical consistency \cite{bd12b}  from bulk cylindrical consistency.

Thus we see that indeed the renormalization trajectory $A^a$ is replaced by the cylindrical consistent set of amplitude maps ${\cal A}_\alpha$. Still one should avoid to equate the scale $a$ with the boundary coarseness $\alpha$.  To consider amplitudes at a certain scale one would have to fix properties of the boundary wave function $\psi_\alpha$ or alternatively for the amplitude kernels ${\cal A}_B(X)$ consider variables $X$ restricted to describe a certain scale. 

The question whether a continuum (or refinement) limit of a quantum gravity model exist can be now reformulated as follows: Does there exist a family of cylindrical consistent amplitude maps that would display the correct semi--classical limit, at least for boundary fields describing a slowly varying geometry or alternatively small curvature. Assuming that slowly varying geometry can be described on a coarse boundary one would need in particular to check the semi--classical limit for simple building blocks ${\cal A}_B$ with a coarse boundary $\alpha$. The semi--classical limit involves to consider a scaling of geometric variables so that these describe lengths much larger than Planck length $l_B(X_{\alpha(B)})>>l_{Planck}$. (Here $l_B$ can be understood as the scale on which the boundary geometry described by $X_{\alpha(B)}$  can vary.) In this limit we expect
\ba\label{semi}
{\cal A}_B (X_{\alpha(B)}) \,\,\sim\,\, \cos( S_H(X_{\alpha(B)}))  
\ea
where $S_H$ is Hamilton's principal function, i.e.\ the action evaluated on the solution determined by the boundary values $X_{\alpha(B)})$. Here we assumed that building blocks will contribute with both possible orientations, as is the case in spin foams. Condition (\ref{semi}) is indeed satisfied for spin foams \cite{asympt}, at least for the simplest building blocks, that is simplices. 

Thus the semi--classicality requirement for the amplitudes is at `mesoscopic' scales $l_B>>l_{Planck}$.  Indeed we need to regularizes the path integral via a discretization. Even classically (non--perfect)  discretization are only reliable reproducing observables which are (much) coarser than the (coarseness) scale of the discretization. If we consider a fixed boundary geometry we can translate this statement into the discretization reproducing observables on scales (much) larger than the discretization scale.

As the cylindrical consistency conditions are very restricting, we can  hope that the condition of cylindrical consistency leads to a unique family of amplitudes, that then define the theory at all scales (i.e. for all boundary wave functions). This philosophy is similar to the asymptotic safety scenario \cite{safe} where one hopes to extrapolate to the UV starting from the IR dynamics of a given theory. Thus the question whether a  refinement limit exist is similar to the asymptotic safety conjecture, namely the existence of an interacting UV fixed point. The question whether we find a unique family of cylindrically consistent amplitudes is connected to the number of relevant couplings at this fixed point, which the asymptotic safety scenario conjectures to be finite. 

This question -- whether a family of cylindrically consistent amplitudes exist or not -- will also determine the allowed matter couplings. The reconstruction of the renormalization flow in terms of the usual notion of scale, as discussed further below, should also reproduce the flow of the standard model matter couplings -- as far as known. Thus including matter couplings would also mean to construct an UV completion of the corresponding quantum field theories -- if such UV completions exist.  One expects restrictions on the allowed matter content - as has been already shown in the asymptotic safety scenario to arise \cite{Astrid}.

As laid out in the previous section, the cylindrical consistent family of amplitudes, that is the renormalization trajectory, can be constructed via an iterative coarse graining procedure. 
The initial amplitudes for this procedure, can be constructed by using a discretization - as is done in the spin foam approach.   The iterative coarse graining procedure reconstructs the renormalization flow in a larger and larger space of `couplings', that also include the parametrization of discretization ambiguities. 
 With respect to the auxiliary coarse graining  flow, that is used to construct the family of cylindrically consistent amplitudes, one can apply the usual notions of relevant / irrelevant couplings and universality. Thus discretization ambiguities (irrelevant couplings with respect to this flow) are taking care off,  see \cite{steinhaus11} for an explicit  example. This addresses the issue (c) in section \ref{dynamics}. Of course one would hope that the flow does not change the semi--classical property (\ref{semi}) of the initial amplitudes, i.e. that the integrated out quantum effects do not change the amplitudes at mesoscopic scales in the sense described above.

A notion of flow, nearer to the Wilsonian one based on scale, would require a reconstruction of this scale from the geometric boundary data. For this one needs to find a way to decompose the geometric variables into small and large scale ones and to correspondingly organize the amplitudes into families of effective ones by integrating out small scale degrees of freedom. This procedure would basically involve the continuums amplitudes encoded in the cylindrical consistent family.  A problem is then to find a (preferable non--perturbative) notion of geometric scale.  Of course also with respect to this flow one can classify relevant / irrelevant couplings, which are now nearer to the standard notion.


\section{ (Decorated) Tensor network renormalization for spin nets and spin foams}\label{dtnws}

The construction of cylindrical consistent amplitude maps is a highly demanding task -- it basically requires to solve the theory for arbitrary complicated boundary data. One  rather hopes for an efficient approximation scheme. The parameter describing the approximations is naturally given by the coarseness $\alpha$.  We can understand this parameter to determine the complexity of boundary data. This approximation scheme is similar to the calculation of  scattering amplitudes for more and more particles (at infinity). Similar to the expectation that for a scattering amplitude involving few particles at infinity in--between states with many particles are less relevant, one can hope that the coarser the boundary data the less relevant become in--between states involving very fine $\alpha$. For this to hold true it is essential that the embedding maps -- that determine the properties of excitations supported by the discrete structure $\alpha$,  are derived from the dynamics of the system. 

Tensor network coarse graining schemes \cite{tnw} implement a recursive improvement of the amplitudes as in (\ref{imp}) and (\ref{eff}). The name `tensor network' indicates that the amplitudes are encoded into tensors associated to vertices (and dual to space time regions). The indices of a tensor at a vertex $v$ are associated to edges attached to the vertex $v$. These edges are also dual to the boundary of the space time region (i.e. the edges cross the boundary). Gluing two space time regions is then equivalent to contracting two indices of two neighbouring tensors.

The complexity of the boundary data, that is the coarseness parameter $\alpha$ translates here into the rank of the tensor and the index range, the so--called bond dimension $\chi$ (assuming finite dimensional Hilbert spaces, which are associated to the edges). Note that  several edges (indices) of a tensor can be summarized into one edge (index) -- thus the bond dimension might increase during the algorithm.

Let us explain an algorithm for a 2D model, in which the amplitudes are encoded into rank four tensors with bond dimension $\chi$. Thus we discretize the partition function (or path integral) with a regular square lattice, where the squares are dual to the four--valent vertices.

One now glues four of such squares to a new  square. This however also increases the number of edges, i.e. the amount of boundary data -- the bond dimension is now $\chi^2$. One needs to reduce these back to the original size $\chi$ (which can be chosen to be much larger than the index range of the original tensors). 

\begin{figure}
\begin{center}
\begin{tikzpicture}[scale=0.3]
\draw [thick] (0,0) rectangle (3,3)
      (5,0) rectangle (8,3);
\draw (1.5,1.5) node {{\Large $M$}}
      (6.5,1.5) node {{\Large $M$}};
\draw (3,1) -- (5,1)
      (3,2) -- (5,2)
      (0,2.5) -- (-1,2.5)
      (0,2) -- (-1,2)
      (-0.5,1.5) node {{ $\vdots$}}
      (0,0.5) -- (-1,0.5)
      (-2,1.5) node {\Large $A$}
      (8,2.5) -- (9,2.5)
      (8,2) -- (9,2)
      (8.5,1.5) node {{ $\vdots$}}
      (8,0.5) -- (9,0.5)
      (10,1.5) node {\Large $B$}
      (4,0.6) node {{\large $\beta$}}
      (4,2.25) node {{\large $\alpha$}};
\end{tikzpicture}  \quad \quad 
\begin{tikzpicture}[scale=0.5]
\draw [thick] (0,0) rectangle (3,3);
\draw (1.5,1.5) node {{\Large $M$}};
\draw (3,1) -- (5,1) arc(-90:90:0.5)
      (3,2) -- (5,2)
      (5.5,1.5) -- (7,1.5)
      (5.5,1.5) node {\large \textbullet}
      (7,1.75) node {\Large $i$}
      (5.9,0.9) node {\Large $V$}
      (0,2.5) -- (-1,2.5)
      (0,2) -- (-1,2)
      (-0.5,1.5) node {{ $\vdots$}}
      (0,0.5) -- (-1,0.5)
      (-2,1.5) node {\Large $A$}
      (4,0.6) node {{\large $\beta$}}
      (4,2.25) node {{\large $\alpha$}};
\end{tikzpicture}
\caption{Left: Two vertices in a tensor network, encoded in the matrices $M$, are sharing two edges with labels $\{\alpha,\beta\}$, which have a total range of $\chi^2$. Right: From the singular value decomposition we can define the map $V$ depicted as a three--valent vertex, where we restrict the label $i$ of the singular values to be $\leq \chi$. \label{svdfig}}  
\end{center}
\end{figure}
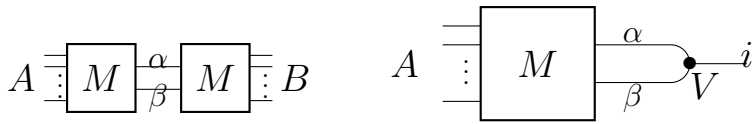

In the case of tensor network methods one chooses a truncation -- via an embedding map as in (\ref{eff}) -- that is chosen from the dynamics of the system. The idea is to approximate as well as possible the summation between two tensors. The situation is depicted in figure \ref{svdfig}. One summarizes the indices of the tensors such that we can rewrite them into matrices $M$. 
We would like to replace the edges carrying an index pair $\{\alpha,\beta\}$ of size $\chi^2$ with an effective edge carrying only a number $\chi$ of indices.  An optimal truncation for the summation over the  index pair $\{\alpha,\beta\}$,   is given by the singular value decomposition of $M_{A\alpha\beta}$:
\ba\label{svd}
M_{A\alpha\beta} = \sum_{i=1}^{\chi^2} U_{Ai} \lambda_i V_{i \, \alpha\beta}
\ea 
where $\lambda_1\geq \lambda_2\geq\ldots\geq \lambda_{\chi^2}\geq 0$ are positive, and $U,V$ are unitary matrices. The truncation drops the smaller set of singular values $\lambda_i$ with $i > \chi$.   Pictorially $V_{i\alpha\beta}$ restricted to $i\leq \chi$ defines a three--valent vertex and we can use these three--valent vertices as in figure \ref{svdfig} to arrive at a coarse grained region with less  boundary data.

Applying the three--valent tensors to the square just glued, we obtain a new effective tensor, with the same bond dimension as before. This algorithm is applicable to systems with and without a background scale. For both cases one hopes that the truncation picks indeed the coarse (homogeneous) data. 

This is supported by several examples, see the discussion in \cite{timeevol}. A better truncation could be reached by choosing the embedding maps to be more non--local (i.e. involving all boundary data and not only those associated to a pair of edges). Indeed in this case the truncation can be even made exact \cite{timeevol}.  To see this consider  a `radial' evolution from a coarser to a finer boundary. The evolution operator only maps to a subspace of the  target Hilbert space with dimension equal or smaller than the initial Hilbert space. Thus a singular value decomposition would turn out to have only as many non--vanishing singular values as we would take into account in the truncation.
A certain notion of locality is however needed to be able to glue the new squares to each other. For a more non--local truncation scheme than the one described here see \cite{second}. 

Such 2D algorithms have been successfully applied to spin net models \cite{finite}, which are analogue models  to spin foams, that can be also defined in 2D \cite{eckert,qgroup} . The spin nets can also be interpreted as specific (`melonic') spin foams, see \cite{qgroup} and are conjectured to have similar statistical properties to spin foams. To be able to do numerical simulations the models considered so far are based on either finite groups or quantum groups $SU(2)_k$. The latter are conjectured to describe quantum gravity with a cosmological constant \cite{smolinetal}. 

The group symmetry protecting variant of the algorithm developed in \cite{eckert,qgroup} allows to keep track of the behaviour of intertwiner degrees of freedom, which signify the status of the simplicity constraints -- the ingredient of spin foams that distinguishes them from standard lattice gauge theories. The initial model differ in the choice of these simplicity constraints. This allows to scan an entire set of models for a reasonable continuum limit. To this end one needs to find a good parametrization of the initial phase space \cite{hol1,wojtek,qgroup}. 

 In fact the simplicity constraints lead to a large extension of the phase space of the latter. A very rich structure of topological fixed points (corresponding to phases in statistical model language) and phase transitions (candidates for interacting theories) has been found in \cite{qgroup}, based on a parametrization of intertwiners developed in \cite{wojtek}. The 2D models also allow to study the concepts discussed in section \ref{dynamics}. In particular the notion of dynamical embedding maps and related vacua states describe condensation phenomena -- in the 2D intertwiner models of anyons described by $SU(2)_k$ fusion modules \cite{wojtek,bais}. 
 
 Recently, $SU(2)_k \times SU(2)_k$ spin net models which impose Barrett--Crane \cite{bc} simplicity constraints have been tested \cite{cameron} and show also an interesting phase structure, which arises by only varying the so--called face weights of the model. 

The richness of the phase structure found so far reinforces the hope that spin foams lead to a reasonable continuum limit. Of course one needs to confirm this hope by coarse graining actual spin foam models. These models are more general in their structure then tensor models, which are basically vertex models, with variables on edges and weight on vertices. In spin foam models  variables  do also appear on two--dimensional objects, i.e.\ plaquettes.

Decorated tensor networks \cite{decorated} can deal with this issue in an effective way. Here one returns to representing the partition function as a gluing of building blocks. These building blocks carry boundary variables as prescribed by the initial model in question. A interesting feature of the procedure is that the type of these initial boundary variables is not changed. This allows a much more straightforward interpretation of the coarse graining flow by keeping track of the behaviour of these variables. For spin foams these variables  coincide with the intertwiner degrees of freedom so important for spin nets -- which is  one reason to expect similar behaviour under coarse graining. 
The geometric interpretation of the (spin) variables in spin foams allows to access whether the coarse graining leads in fact to a geometric coarse graining of the system. This feature will in particular be encoded in the embedding maps. 

In lowest order approximation the building blocks will carry (almost) the same amount of boundary data, as the initial model. As mentioned this allows for a straightforward interpretation of the coarse graining flow of these systems. Going to higher order truncations one incorporates more boundary data by associating a tensor to the building blocks which now introduces `higher order' variables. The entire coarse graining procedure is similar to tensor network algorithms (i.e. also based on singular value decompositions), but `decorated' by the original variables of the model. 

Another feature of decorated tensor networks is that they may allow for (semi--) analytical calculations, see also \cite{jeff}. This is important to be able to treat spin foam models based on Lie groups, where  the issue of divergencies arise \cite{divergence,bubble}, see also the discussion in section \ref{diff}.

In tensor network algorithms the truncation is determined by the dynamics. This is so far the only way to find a reliable truncation, but makes the algorithms computationally very demanding. An alternative might be to use truncations, informed by some geometric intuition. E.g. \cite{cuboids} imposes a restriction to discretizations built out of cuboids, that describe geometries without curvature but with torsions. The truncations can be again imposed by an embedding map, that this time is however chosen by hand. The flatness makes the action contribution to the amplitudes vanish, thus the coarse graining flow tests only the measure terms.  This flow does however indicate a restoration of (a remnant of) diffeomorphism symmetry, as we will explain in the next section.

\section{Diffeomorphism symmetry in the discrete, constraints and divergencies} \label{diff}

In this section we are going to elaborate more on a notion of diffeomorphism symmetry in the discrete. This symmetry is a very powerful one \cite{ditt,bd12a}, in fact its realization signifies that the continuum limit has been reached. This is meant in the following way: although the physics is expressed on a discrete structure, the predictions for observables, which can be supported by this discrete structure coincide with those of the continuum model. Such a discretization is called perfect \cite{hasenfratz,bahrdittrich09b} -- it exactly mirrors continuum physics.

Thus the refinement limit is necessary to reach diffeomorphism symmetry and with this a notion of physical states.

The notion of diffeomorphism symmetry we are going to discuss here also arises for discretizations which do not explicitly involve coordinates. For instance in Regge calculus \cite{regge} the variables are given by the lengths of the edges in a triangulation.\footnote{ Alternatively one can use areas and angles \cite{angle}, which is nearer to the variables used in spin foams.} These geometric data of the discrete elements allow to determine the relative position of the vertices with respect to each other.

In fact if there is a symmetry of the action\footnote{That is the Hessian of the action evaluated on a solution needs to have null modes making the solutions non--unique \cite{dittrichreview,bahrdittrich09a}. The action itself (away from solutions) will allow a huge class of invariant deformations, most of these trivialize however if restricted to solutions.} allowing for these relative vertex positions -- expressed in the geometric data of the discretization -- to change, we speak of a realization of diffeomorphism symmetry in the discrete. This symmetry is also referred to as vertex translations, as it coincides in the 3D BF formulation of gravity with the shift or translation symmetry of the triad fields \cite{louapre}.

Such a symmetry has been indeed identified for linearized Regge calculus \cite{rocek} and a number of examples \cite{steinhaus11,bd12a,bahrdittrich09b}. It is however broken  if one considers a (Regge) solution of 4D gravity with curvature \cite{bahrdittrich09a} or perturbative Regge gravity beyond linear order \cite{hoehn1}. Here `broken symmetry' means that the Hessian, instead of null modes, will display modes with very small eigenvalues (compared to the other eigenvalues). This breaking has severe repercussions. It  prevents the path integral -- for the regularization of which we need to introduce the discretization -- from acting as a projector onto physical states. 

One can define a canonical discrete time formulation, consistent\footnote{i.e. reproducing the equations of the covariant framework} with the covariant one \cite{consistent,hoehn1}. This formalism transfers also consistently the (broken) symmetries into (pseudo) constraints. Whereas constraints are given as equations of motions that only involve the canonical data of one time step, pseudo constraints will also involve, with a weak dependence, the data of a neighbouring time step.  

Thus one reason to take the refinement limit  is actually to restore the diffeomorphism symmetry \cite{nielsen,bahrdittrich09b}, as is also used in the perfect action program \cite{hasenfratz}  for lattice QCD with regard to Lorentz symmetry. There are a number of arguments for such a restoration: one is that the pseudo gauge modes should have a small  lattice correlation lengths  and decouple in the continuum limit \cite{nielsen}. Another is that for instance for Regge calculus with flat building blocks the eigenvalues of the Hessian of the action associated to the pseudo gauge modes scale with the curvature per building block of the solution \cite{bahrdittrich09a}. In the refinement limit this curvature goes to zero, thus leading to a restoration of the symmetries. 

We described the symmetry as allowing displacements of vertices. This is basically the reason why this symmetry is so powerful and requires the continuum limit: For a system with such a symmetry it means that, given a solution, we can move the vertices around (i.e. change the associated geometric data) without changing the value of the action. Thus we can for instance move vertices on top of each other, reaching a coarser triangulation. This is basically the argument that diffeomorphism symmetry implies triangulation independence \cite{steinhaus11}.  One can furthermore move the vertices such that one region appears very finely grained and another very coarse grained. (Again this is with respect to a solution, which provides a scale). Thus our model has to display continuum physics reliable on all scales and show no discretization artifacts, i.e. it has to be a perfect discretization. Such a perfect discretization avoids all problems (ambiguities, breaking of symmetries, triangulation dependence) of a `typical' discretization.

For interacting systems one can hope only for non--local actions or amplitudes to display such a powerful symmetry. This is shown explicitly, with the non--existence of a local path integral measure for linear Regge calculus \cite{measure}. Non--local amplitudes are very difficult to deal with -- in fact the framework described in sections \ref{dynamics} and \ref{dtnws} avoids non--local couplings by introducing building blocks with more boundary data -- akin to introducing more fields in the continuum to absorb higher derivatives. 
Since diffeomorphism symmetry implies triangulation invariance we can also hope that coarse graining schemes on a regular lattice are sufficient to recover fully triangulation invariant models, which is indeed confirmed so far for spin net models \cite{wojtek,eckert,qgroup}. 

As noted above (first class Hamiltonian and diffeomorphism) constraints can only appear if the discretization shows diffeomorphism symmetry and hence is perfect. Thus for 4D gravity one has to expect non--local constraints. Again, the framework developed in section \ref{dynamics} could be of help here, as it might be possible to derive constraints on very coarse Hilbert spaces ${\cal H}_\alpha$ first and then going to finer and finer ones. This does not exclude graph--changing Hamiltonians \cite{thiemannH}, although one would expect that an inductive Hilbert space based on dynamical embedding maps, allows for graph--non--changing ones. In fact for the simplest triangulations, leading to only flat bulk solutions, it is possible to find first class constraints \cite{bonzomdittrich}.  Note that constraints which are derived from cylindrically consistent amplitudes, do also describe the flow of  (matter) coupling constants. This can for example appear in the form of couplings, that depend on the geometric variables associated to building blocks.  This information on the couplings of the running is dynamical information -- which if the constraints are indeed derived from the consistent amplitudes, is obtained from the coarse graining process that led to these amplitudes. It seems  impossible to construct consistent Hamiltonian constraints, without having such an explicit process that determines the running of the couplings.

One can also turn the argument around and say that if a refinement limit does not lead to a restoration of the symmetries (or first class property of the constraints), the system is inherently discrete \cite{gp}. The question of whether a refinement limit `exist' or not might however depend on many details of how the system is constructed as well as how one attempts to construct\footnote{\cite{gp}  makes a choice of (local) constraints, turns these into a master constraint \cite{master,consistent} and tests this master constraint on a certain class of semi--classical states. Each step involves a number of ambiguities.} the continuum limit.  

Let as also remark on the relation between divergencies and diffeomorphism symmetry. As the gauge orbits of this symmetry are non--compact (with the exception of Euclidean gravity with positive cosmological constant) one has to expect that the partition function diverges in the case the symmetries are realized. (Vertex translations may also cross building blocks, reversing orientations of these \cite{orient}, which allows for non--compact orbits.) Thus one would expect a divergence of $\Lambda^{ND}$ for $D$ space time dimension and $N$ triangulation vertices. This is indeed confirmed for (topological) 3D spin foams with the link to the diffeomorphism symmetry made explicit \cite{louapre}. The divergence structure of the 4D models is less clear \cite{divergence,bubble} , as it also depends on a choice of path integral measure in the form of so called edge and face weights \cite{martinalex,bubble}. A correct divergence structure in itself would of course not be sufficient for a model to display diffeomorphism symmetry, as this structure can be easily tuned by only changing face and edge weights \cite{bubble}, but leaving the (discretized) action unaffected. Additionally the existence of degenerate configurations, which may display enhanced symmetries \cite{bubble} and divergencies  complicate the issue.

As symmetries are typically broken one would expect the initial model to be finite. (As noted in \cite{bubble} special configurations might actually exist, which show enhanced symmetries and might lead to divergencies.) Under coarse graining, with the restoration of symmetries, the path integral becomes however more and more divergent. One could expect a problem here, however one can indeed deal with this successfully even in a numerical approach \cite{steinhaus11}. In fact the coarse graining procedure involves a rescaling of the amplitudes in each step. One would then expect that the (candidate) divergencies lead to an enhancement of the terms in the amplitude that do lead to diffeomorphism symmetry and a suppression of the other terms. 
Thus diffeomorphism symmetry might enhance its own restoration in the refinement limit in this way.

\section{Summary and Outlook}\label{diss}

A refinement limit is inescapable for the construction of the full theory of (loop) quantum gravity. 
Only in this limit can we expect the realization of diffeomorphism symmetry, thus a notion of Hamiltonian and diffeomorphism constraints and finally physical states. Indeed as we explained here constructing the refinement limit means to construct physical states and a physical Hilbert space. 

We presented a framework to formulate and construct the refinement limit, using the essential structure of inductive limit Hilbert spaces and the concept of cylindrical consistent amplitudes, where the notion of cylindrical consistency is induced from the dynamics. The (tensor network) coarse graining procedures discussed   construct such amplitudes iteratively in a truncation scheme. The dynamics automatically determines this truncation, by introducing a notion of coarse states with few excitations and very fine states with many excitations. The excitations are with respect to a  vacuum state that is also determined from the dynamics.

Although the construction of the refinement limit requires basically the solution of the model, it can be organized in a truncation scheme. The approximation improves with finer and finer discretizations  taking into account, that support more and more complicated boundary data.  Cosmology rather involves coarse data, one might therefore hope that a derivation of cosmology can be obtained at low truncation order \cite{cosmo} (but sufficiently fine to determine the dynamical embedding maps essential for the understanding of the truncation). 

We laid out the relation of the refinement scheme to renormalization involving a (background) scale. The scale is basically replaced by the coarseness of the discretizations -- although one should be careful in equating the two. The notion of a complete renormalization trajectory  is replaced by the notion of cylindrically consistent amplitudes, showing the correct semi--classical limit behaviour (i.e. for large geometries). The crucial question is whether such cylindrically consistent amplitudes exist.

Renormalization, also in the sense of regulating divergencies, come also up in group field theories (gft's) \cite{oritireview,lreview}. In this case one sums over triangulations and hopes to achieve a continuum limit by choosing weights such that configurations with infinitely many building blocks dominate \cite{melonicphase}. The relation between renormalization in a gft sense\cite{sylvain}, which involves an explicit scale, and the coarse graining scheme presented here should be better understood, in particular since the divergencies (may) correspond to gauge symmetries in the spin foam framework. (A gft understanding of vertex translations leads to global symmetries \cite{gftsymm}.)  

One can argue that due to the restoration of diffeomorphism symmetry in the form of vertex displacement invariance a given sufficiently fine lattice may simulate many coarser lattices. A variant of this argument is used in \cite{smerlak} , to show that refinement and summing over triangulations should lead to the same result.\footnote{See also \cite{lreview,zipfel} for a related discussion of this issue, namely whether the sum over triangulations leads to the path integral as projector.} To inquire more about this relation, it is essential to clarify the relations between the Hilbert spaces involved, as the notion of cylindrical consistency is rather different \cite{oritiLQG}. Indeed, whether one prefers refining or summing over triangulations to obtain (bulk) triangulation independent amplitudes, in both cases we demand the amplitudes to be cylindrically consistent with respect to some choice of embedding maps. This latter notion specifies the relation between different boundary discretizations, and turns the amplitudes into well--defined maps on the continuum Hilbert space.

Let us comment on some possible future developments. Coarse graining results from spin net models hint at a rich phase space structure for spin foams. With an explicit coarse graining scheme for spin foams at hand \cite{decorated} we can expect  results to arrive soon, that will allow deep insights into the dynamical mechanisms of spin foams -- and thus hopefully the workings of quantum spacetime.

Even the identification of topological field theories in the phase diagram of spin foams can give rise to exciting developments. Such topological field theories lead to cylindrical consistent embedding maps, thus to new inductive limit Hilbert spaces \cite{timeevol,newvac,DG16}.   These can be used to construct  further alternative vacua and representations for loop quantum gravity\footnote{The uniqueness results \cite{lost} pertaining to the Ashtekar Lewandowski representation do not apply as flux operators may exist only in exponentiated form}, possibly with a notion of simplicity constraints and in--between the Ashtekar Lewandowski representation \cite{AL} and the one based on $BF$ theory developed in \cite{newvac}. Different vacua and representations allow to expand the theories around different regimes and to thus organize the dynamics of the theory with respect to different notions of excitations. This opens  new perspectives for loop quantum gravity and can lead to a large extension of the framework.

\section*{Acknowledgements}
This research was supported by Perimeter Institute for Theoretical Physics.
Research at Perimeter Institute is supported by the Government of Canada through Industry Canada and by the
Province of Ontario through the Ministry of Research and Innovation.

\end{document}